\documentclass[twocolumn]{aastex61}
\usepackage{graphicx}
\usepackage{natbib}
\usepackage{amsmath}
\usepackage{dsfont}
\usepackage{amssymb}
\usepackage{algorithm} 
\usepackage[noend]{algpseudocode}
\usepackage{varwidth}
\bibstyle{apj}
\bibpunct{(}{)}{;}{a}{}{,}
\usepackage{txfonts}
\usepackage{float} %figure inside minipage
\def\tens#1{\ensuremath{\mathsf{#1}}}
\def\BF#1{{#1}}
\newcommand\bigzero{\makebox(0,0){\text{\huge0}}}

\newcommand{\A}           {{\tens{A}}}
\newcommand{\F}           {{\tens{F}}}

\newcommand{\E}           {{\tens{E}}}

\newcommand{\J}           {{\tens{J}}}

\newcommand{\ostar}       {{\circledast}}
\newcommand{\AWP}         {{\it A-Projection}}
\newcommand{\myindent}[1]{
\newline\makebox[#1cm]{}
}

\def\JS#1#2#3{{\tens{J}^{#2}_{\!#1}#3}}
\def\MS#1#2#3{{\tens{M}^{#2}_{#1}#3}}
\def\E#1#2#3{{\tens{E}^{#2}_{#1}#3}}

\def\BARN#1#2{{\left.{#1}\right|_{#2}}}
\begin{document}
\title{The Pointing Self Calibration algorithm for aperture synthesis radio telescopes} 
% Should say radio: discussion throughout is for radio context
% \title{Direction dependent calibration of aperture synthesis
%   telescopes: The Pointing SelfCal algorithm} 

\shorttitle{The Pointing Selfcal algorithm}
\shortauthors{Bhatnagar and Cornwell}

\correspondingauthor{S. Bhatnagar}
\email{sbhatnag@nrao.edu}
\author{S. Bhatnagar}
 \affiliation{National Radio Astronomy Observatory,
  1003 Lopezville Road, Socorro, NM, 87801, USA.}
\author{T.J.~Cornwell} 
 \affiliation{National Radio Astronomy Observatory,
  1003 Lopezville Road, Socorro, NM, 87801, USA.}
\affiliation{Tim Cornwell Consulting}%\email{realtimcornwell@gmail.com}

\date{Received: June 13, 2017;  Accepted: Sept. 22, 2017 }

\begin{abstract}
  This paper is concerned with algorithms for calibration of direction
  dependent effects (DDE) in aperture synthesis radio telescopes
  (ASRT). After correction of Direction Independent Effects (DIE)
  using self-calibration, imaging performance can be limited by the
  imprecise knowledge of the forward gain of the elements in the array.
  In general, the forward gain pattern is directionally dependent and
  varies with time due to a number of reasons. Some factors, such as
  rotation of the primary beam with Parallactic Angle for
  Azimuth-Elevation mount antennas are known {\em a priori}. Some,
  such as antenna pointing errors and structural
  deformation/projection effects for aperture-array elements cannot be
  measured {\em a priori}. Thus, in addition to algorithms to correct
  for DD effects known {\em a priori}, algorithms to solve for DD
  gains are required for high dynamic range imaging.  Here, we discuss
  a mathematical framework for antenna-based DDE calibration
  algorithms and show that this framework leads to computationally
  efficient optimal algorithms which scale well in a parallel
  computing environment.  As an example of an antenna-based DD
  calibration algorithm, we demonstrate the Pointing SelfCal algorithm
  to solve for the antenna pointing errors.  Our analysis show that
  the sensitivity of modern ASRT is sufficient to solve for antenna
  pointing errors and other DD effects.  We also discuss the use of
  the Pointing SelfCal algorithm in real-time calibration systems and
  extensions for antenna Shape SelfCal algorithm for real-time
  tracking and corrections for pointing offsets and changes in antenna
  shape.
\end{abstract}
%\maketitle

%%%%%%%%%%%%%%%%%%%%%%%%%
\section{Introduction}
The scientific deliverables of modern and next generation
interferometric radio telescopes, some under operation or in advanced
stages of commissioning, naturally require high sensitivity and high
dynamic range imaging. All these telescopes typically promise at least
a ten-fold increase in instantaneous sensitivity compared to previous
generation telescopes.  The projected achievable thermal noise in the
images from these telescopes is in the range of $1-10$~$\mu$Jy/beam
corresponding to typical imaging dynamic ranges of $10^{5-7}$,
particularly at frequencies $<$few~GHz.

The underlying assumption in sensitivity calculations is that the data
processing procedure will remove systematic effects due to the
instrument, atmosphere/ionosphere or sky to a level significantly
below the thermal noise limit so that the RMS noise decreases as
square root of the total number of independent measurements (the
product of the total observing bandwidth ($\Delta \nu$) and total
on-source integration time ($\Delta T$)).  In practice, the data are
corrupted due to a number of direction independent (DIE) and direction
dependent (DDE) effects.  DI effects are constant across the
field-of-view (FoV) and correspond to a single value per antenna as a
function of time, frequency, and polarisation.  Efficient algorithms
to solve for DI effects have been in use for many decades
\citep{THOMPSON_AND_MORAN}.  However some instrumental and
atmospheric/ionospheric effects are directionally dependent.
Designing efficient solvers for DDE is more difficult and lag behind
solvers for DIE.

Aperture synthesis radio telescopes synthesize an aperture equivalent
to the largest projected separation between the individual antennas in
the array by cross-correlating the signals from each antenna-pair in
the array \citep{THOMPSON_AND_MORAN}.  Before being correlated,
imprinted on the signals are the effects of a number of instrumental
and atmospheric/ionospheric
components.  The far-field complex-valued electric field pattern (EFP) of the
antenna and the feed further modifies the incoming wave before it is
detected as a voltage fluctuation.  In general, the cumulative effect of 
all these is time-, frequency-, polarization- and direction-dependent.

The EFP of the individual elements in an aperture synthesis array
constitutes the strongest instrumental DD effect.  While it is
fundamentally direction-, frequency- and polarization- dependent, it
can also vary significantly with time for a number of reasons.
The primary beam (PB) is typically rotationally {\it asymmetric} and
for 2-axis Azimuth-Elevation mount (Az-El mount) dish antennas, it
rotates with respect to the sky as a function of the Parallactic Angle
(PA).  This leads to time varying DDE gains within the field-of-view
(FoV) and constitutes the strongest time varying DDE for such systems.
For aperture-array elements, the beam geometry is fixed in Earth-based
coordinates, and thus varies in celestial coordinates. This
constitutes the strongest instrumental DD effect for
aperture-array telescopes.  Given an antenna aperture illumination
function measured (or modeled) {\it a priori}, the existing \AWP\
algorithm can be used to correct for these known effects during image
deconvolution \citep{AWProjection,WB-AWP}.

Antenna pointing errors affect the approximately static case of
snapshot imaging with telescopes with high instantaneous sensitivity.
In observations with long integration time (for improved sensitivity
or uv-coverage, or both), time varying antenna pointing offsets also
lead to time-varying DDE gains comparable in magnitude to those due to
the rotation of PB with Parallactic angle (PA).  In this paper, we
describe a general mathematical framework for algorithms to solve for
the {\it unknown} DDE. We apply it to the specific case of antenna
pointing errors.

\section{Theoretical framework}
\label{Sec:Theory}

The measurement equation including DDE for an aperture
synthesis telescopes can be compactly written using the
notation in \cite{HBS1} as\footnote{The symbol '$\iota$' is used
  as a symbol for iota throughout the text and should not be confused with $i$
  which is used as an antenna-index subscript.}
\begin{eqnarray}
\label{Eq:ME}
V^{Obs}_{ij} &=& \MS{ij}{DI}{}\int \MS{ij}{DD}{}(\vec{s})~I(\vec{s})~
e^{2\pi\iota \vec{s}\cdot\vec{b_{ij}}}~d\vec{s}\nonumber\\
&=& \MS{ij}{DI}{} \left(\A_{ij}\star V^\circ \right).
\end{eqnarray}
The appropriate symbols here and in the rest of the paper represent
quantities in the instrumental polarization basis (circular or linear
polarization).  $V^{Obs}_{ij}$ is the observed and $V^\circ$ the true
full polarization visibility vectors. $\A_{ij}$ is the Fourier
transform of $\MS{ij}{DD}{}$ and the symbol '$\star$' represents the
standard matrix-vector multiplication, except that the multiplication
operations are replaced by the convolution operation in the algebra.
$\MS{ij}{DI}{}$ and $\MS{ij}{DD}{}(\vec{s})$ are the outer products of
the DI and DD Jones matrices $\JS{}{DI}{}$ and $\JS{}{DD}{(\vec{s})}$
respectively.  We refer to these outer products as the {\it Radio}
Mueller matrices\footnote{In the original optical literature
  \citep{Jones, Mueller}, Jones matrices are defined in the Stokes
  basis and an outer product of the Jones matrices is called the
  Mueller Matrix.  Radio interferometric measurements are an outer
  product of electric fields measured at the two antennas in the feed
  polarization basis (circular or linear) which is the natural basis
  for calibration.  However since this is also specific to radio
  interferometric measurements, we use the term {\it Radio Mueller
    matrix}.  This Radio Mueller matrix is related to the optical
  {Mueller} matrix via a unitary transform \citep[see
  e.g.,][]{HBS1}.
} for DI and DD
gains.  $\vec{s}$ is a direction in the sky, $I$ is the
image and $\vec{b_{ij}}$ is the projected separation between the
antennas $i$ and $j$ in units of the wavelength.  The goal of imaging
is to estimate the {\it true} sky brightness distribution $I(\vec{s})$
in the presence of known or unknown gains $\MS{ij}{DI}{}$ and
$\MS{ij}{DD}{}(\vec{s})$.

\subsection{Overview of direction independent calibration}

Direction independent effects (DIE) ($\MS{ij}{DI}{}$ in
Eq.~\ref{Eq:ME}) are due to the telescope electronics and
atmospheric/ionospheric effects at scales much large than the antenna
FoV.  Calibration of such gains is done using the self-calibration
technique \citep{SELFCAL_CORNWELL,SELFCAL_LECTURE}.  Ignoring
$\MS{ij}{DD}{(\vec{s})}$ and factoring $\MS{ij}{DI}{}$ into the
antenna based DIE $\JS{i}{}{}$, a convenient specialization of
Eq.~\ref{Eq:ME} for DIE calibration can be written as
\begin{equation}
\label{Eq:MEDI}
V^{Obs}_{ij} = \left(\JS{i}{}{}\otimes\JS{j}{*}{}\right) V^\circ
\end{equation}
and $\JS{i}{}{}$ solved-for using $\chi^2-$minimization techniques.
For model visibilities $V^M_{ij}$, the residual visibilities are
\begin{equation}
\label{Eq:Res}
R_{ij}=V^{Obs}_{ij}-( \JS{i}{}{} \otimes \JS{j}{*}{})V^M_{ij}
\end{equation} 
and 
\begin{equation}
\label{Eq:Chisq}
\chi^2=\sum_{ij}R_{ij}^{\dag}\cdot W_{ij}\cdot R_{ij}.  
\end{equation}
\BF{$W_{ij}$ is a diagonal matrix of weights
proportional to the inverse of the measurement variance.}
Equation \ref{Eq:Chisq} is a sum over all baselines of the
weighted L2-norm of the full-polarization residual vector $R_{ij}$.  For
clarity, we show below the expansion of Eq.~\ref{Eq:Chisq} in terms
of the elements of $R_{ij}$ and $W_{ij}$:
\begin{eqnarray}
\chi^2 &=& \sum_{ij}
\left\{
%\left[
\begin{bmatrix} 
r^{pp^*}~r^{pq^*}~r^{qp^*}~r^{qq^*}&
\end{bmatrix}_{ij}
\begin{bmatrix}
    w^{pp} &        &       & \bigzero~~~ &  \\ 
          & w^{pq}  &       &          &  \\ 
~~~\bigzero  &        & w^{qp} &          &  \\ 
          &        &       & w^{qq}    &  \\ 
\end{bmatrix}_{ij}
\begin{bmatrix}
r^{pp}\\
r^{pq}\\
r^{qp}\\
r^{qq}
\end{bmatrix}_{ij}\right\}\nonumber\\
&=& \sum_{ij} \left( \left| r^{pp}\right|^2 w^{pp} + \left|
    r^{pq}\right|^2 w^{pq} + \left| r^{qp}\right|^2 w^{qp} + \left| r^{qq}\right|^2 w^{qq}\right)_{ij}
\end{eqnarray}
where $r$ and $w$ are the elements of the vector $R_{ij}$ and the
matrix $W_{ij}$ respectively and subscripts $p$ and $q$ represent
orthogonal polarization states (circular or linear).

The Jones matrix $\JS{i}{}{}$ in Eq.~\ref{Eq:Res} are direction
independent with its elements consisting of single numbers (and not 2D
functions). The evaluation of $\MS{ij}{DI}{}$ in Eq.~\ref{Eq:ME}
therefore requires computationally simpler outer-product operator (as
against the outer-convolution operator required for $\MS{ij}{DD}{}$;
see Section~\ref{Sec:DD}) and application of
$\left(\JS{i}{}{}\otimes\JS{j}{*}{}\right)$ is a simple matrix
multiplication.  This observation leads to an efficient DI calibration
algorithm.  \BF{For the relevant case of parallel-hand only calibration,
\begin{equation}
\JS{i}{}{}=\left[
    \begin{array}{cc}
    g^{p} & 0       \\ 
      0  & g^{q}    \\ 
  \end{array}\right]_i
\end{equation}
where the superscripts $p$ and $q$ represent the two orthogonal
polarization pairs. $R_{ij}$ can be written as
\begin{equation}
\label{Eq:ResXij}
R_{ij}=X_{ij}-( \JS{i}{}{} \otimes \JS{j}{*}{})\mathds{\vec{1}}
\end{equation} 
where the symbol $\mathds{\vec{1}}$ is a $4\times 1$ column-vector
with all elements equal to one and}
\begin{equation}
  X_{ij}=\left[diag(V^{M}_{ij})\right]^{-1}V^{Obs}_{ij}.\nonumber
\end{equation}
\BF{The function $diag(\vec{a})$ returns a diagonal matrix with the vector
$\vec{a}$ as its diagonal.} For a simple minimization algorithms such
as the Steepest Descent algorithm, equating the gradient
$\frac{\partial \chi^2}{\partial \JS{i}{*}{}}$ to zero for an
$N_{ant}$ antenna-array leads to $N_{ant}$ simultaneous non-linear
equations which are then solved using the following iterative
equation:
\begin{equation}
\label{Eq:SelfCal}
\begin{aligned}
\BARN{\JS{i}{}{}}{n}\otimes\mathds{1}=&\left(1-\gamma\right)\left[\BARN{\JS{i}{}{}}{n-1}\otimes\mathds{1}\right] + \\
           &\gamma\left[SoW\right]^{-1}\left[ {\sum\limits_{j, j\ne i}\left( \BARN{\JS{j}{}{}}{n-1}\otimes \mathds{1}\right) W_{ij} X_{ij} }\right]
\end{aligned}
\end{equation}
where
\begin{equation}
  SoW={\sum\limits_{j, j\ne i }\left(\BARN{\JS{j}{}{}}{n-1}\otimes \BARN{\JS{j}{*}{}}{n-1}\right) W_{ij}}\nonumber
\end{equation}
The symbol $\BARN{}{n}$ represents value at iteration $n$, $0<\gamma<1$ is
the standard feedback loop-gain of non-linear minimization algorithms
and $\mathds{1}$ is the $4\times4$ identity matrix.
When $V_{ij}^{M}$ is a wide-band prediction of the sky brightness
distribution in the FoV, $X_{ij}$ is a weak function of time and
frequency and it may be pre-averaged for the solution interval for
improved signal-to-noise ratio (SNR).  Using previous solutions or
unity as an initial guess for $\JS{i}{}{}$, convergence is typically
achieved in a few iterations (see
\cite{FREQ_RESPONSE_OF_INTERFEROMETER, RANTSOL}\footnote{Also
  accessible from \\
  http://www.aoc.nrao.edu/$\sim$sbhatnag/GMRT\_Offline/antsol/antsol.html}
for a detailed derivation and an intuitive interpretation).  

\BF{The primary focus of this paper is DD calibration -- described in
  the following sections -- which is fundamentally coupled to
  imaging.  DD calibration is therefore better described in a
  Mueller-matrix framework.  To establish equivalence between
  DI and DD calibration and to show that DD calibration is a
  generalization of the DI calibration, we also described the DI
  calibration above in a Mueller matrix formulation.  In practice, most
  software implementations of Eq.~\ref{Eq:SelfCal} implement the
  matrix arithmetic by-hand for better code optimization where the
  distinction between Jones- and Mueller-matrix based formulations is
  not important.   For {\it
    DI-only} implementations which may benefit from directly using the
  matrix arithmetic, it is more efficient to re-formulate the DI
  algorithm using $V^{Obs}_{ij} = \JS{i}{}{}V^\circ\JS{j}{*}{}$ where
  all matrices are $2\times2$ matrices,
  instead of Eq.~\ref{Eq:MEDI}.  However it is important to point out here
  that while this may offer computational advantages, it does not lead
  to a fundamentally new algorithm.}

Since $V^M$ in Eq.~\ref{Eq:Res} is independent of $\JS{}{}{}$, it is
treated as a constant in the minimization algorithm.  Calibration for
DIE gains and imaging to solve for $I(\vec{s})$ are alternated
iteratively.  Thus calibration can be done keeping the model of the
sky fixed and imaging is done keeping the DI calibration terms fixed.
Iterating between imaging using calibrated data
$V^C_{ij} = ( \JS{i}{}{} \otimes \JS{j}{*}{})^{-1}V^{Obs}_{ij}$ to
make the model image $I^M$ and using $I^M$ to solve for $\JS{i}{}{s}$
forms a closed-loop solution to account for DIE.  We retain this
overall structure for our DDE calibration approach and substitute an
algorithm for estimating the DDE's while keeping the image fixed.

\subsection{Direction dependent effects antenna-based calibration}
\label{Sec:DD}
Direction dependent effects (DDE) are represented by $\MS{ij}{DD}{}$
in Eq.~\ref{Eq:ME}.  Mathematically the effect of these terms is
indistinguishable from $I(\vec{s})$ and the model data $V^M$ cannot be
evaluated independent of $\MS{}{DD}{}$.  Consequently, to compute the
residual vector (the difference between data and the model), the
integral in Eq.~\ref{Eq:ME} must be evaluated for each measured data
point (i.e., for all $i$, $j$, frequency and polarization
measurements).  With typical modern data sizes in the multi-Tera bytes
regime the computational and the data I/O costs become very high even
with only a few unresolved sources for a direct evaluation of the
integral, or evaluating it separately for different directions in the
FoV.  These costs are prohibitive for dense fields (as is typical at
frequencies $\le few$~GHz) and for fields with a combination of
compact and significant extended emission (as is typical for
observations in the Galactic Plane and for mosaic observations at any
frequency).

Direction dependent effects that are fundamentally aperture plane
effects can often be compactly modeled with a few parameters in
the aperture plane.  For example, the effect of antenna pointing errors can be
modelled by two parameters per antenna for the entire FoV.  Solving for these effects in
the image plane requires solving for the antenna gains towards
multiple sources.  This has significant numerical and computational
disadvantages.  On the other hand, solving for these effects directly in the aperture
plane has lower computational complexity and optimal utilization of
the full SNR based on the integrated flux in the FoV.  The Pointing
Selfcal (PSC) algorithm described below is an example of such an
aperture-plane calibration algorithm for antenna pointing errors.

The $\A_{ij}$ in Eq.~\ref{Eq:ME} can be factored into antenna-based
Jones matrices. The observed visibilities $V_{ij}^{Obs}$ calibrated for
$\MS{}{DI}{}$ can then be written as 
\begin{equation}
\label{Eq:MEDD}
V_{ij}^{Obs} = \left[ \A_i\ostar \A_j^{*} \right] \star V^{M^\circ},
\end{equation}
where $V^{M^\circ}=\F~I(\vec{s})$ -- the Fourier transform of the sky
brightness distribution on a grid.  The symbol '$\ostar$', introduced in
\cite{AWProjection}, represents the outer-convolution
operator\footnote{The element-by-element algebra of the
  outer-convolution operator is the same as that of the outer-product
  operator used in the DI description of \cite{HBS1}, except that the
  complex multiplications are replaced by convolutions.}.  $\A_i$ is
an antenna DD Jones matrix given by
\begin{equation}
\label{Eq:AJones}
\A_i = 
\begin{bmatrix}
A^{p} & 0 \\
0    & A^{q} 
\end{bmatrix}_{i}.
\end{equation}
The elements along the diagonal of this matrix describe the electric field distribution across the
antenna aperture for the two orthogonal polarizations.

The gradient of $\chi^2$ w.r.t. antenna-based parameters $a$ (which
are, in general, complex-valued) for
the more general Eq.~\ref{Eq:MEDD} is given by\footnote{Equations~\ref{Eq:DCHIDD}
  and \ref{Eq:DA} are a general form of equations for antenna-based
  calibration and include the DI case.  As a test and for
  intuitive understanding, replacing $\star$ by dot product, $\A_{ij}$
  by $\JS{i}{}{}\otimes\JS{i}{*}{}$, $a^{*}_{\!i}$ by
  $\J^{*}_{\!i}$ and equating Eq.~\ref{Eq:DCHIDD} to zero recovers
  Eq.~\ref{Eq:SelfCal}.}
\begin{equation}
\label{Eq:DCHIDD}
\BARN{\frac{\partial \chi^2}{\partial a^{*}_{\!i}}}{n} = -2 \sum\limits_{j,j \ne i}
\Re \left( \BARN{R_{ij}^{\dag}}{n} \cdot
  W_{ij}\cdot\left[\BARN{\nabla_{\!\!i} V^M_{ij}}{n}\right]\right),
\end{equation}
where
%\begin{eqnarray}
\begin{flalign}
\label{Eq:DA}
&&\nabla_{\!\!i} \A_{ij} &= \frac{\partial \A_{ij}}{\partial a^{*}_{\!i}} = \frac{\partial
      \A^M_i}{\partial a^{*}_{\!i}}\ostar \A^{M^{*}}_{\!j}&\\
\text{and}
&&\BARN{\nabla_{\!\!i} V^M_{ij}}{n} &=\BARN{\nabla_{\!\!i} \A_{ij}}{n} \star  V^{M^\circ}.&\nonumber
\end{flalign}
The symbol $\Re$ represents the real part of its argument, which
evaluates to a scalar in the same way as Eq.~\ref{Eq:Chisq}.  $\A^M_i$
-- the model for the true $\A_i$ -- is parametrized by the parameter
$a$ and
$\BARN{R_{ij}}{n}=V_{ij}^{Obs} - \left[ \A^M_i \ostar
  \A^{M^{*}}_{\!j}\right]_n\star V^{M^\circ}$ is the residual vector
computed at iteration $n$.  The parameters are then updated
iteratively as
\begin{equation}
\BARN{a}{n} = \BARN{a}{n-1} + f\left(\BARN{\frac{\partial \chi^2}{\partial a^{*}}}{n}\right) 
\end{equation}
where $f$ is a function that depends on the details of the non-linear
minimization algorithm used.  For minimization algorithms that assume
a diagonally-dominant Hessian matrix (such as the Steepest descent
algorithm), $f(x)=\gamma x$.  More sophisticated minimization
algorithms involve potentially expensive evaluation of the
Jacobi/covariance matrix \cite[see][or later editions for detailed
discussions]{NumericalRecipes}.

\section{The Pointing Selfcal algorithm}

Our algorithm works in the general way of self-calibration algorithms,
iterating between estimation of the sky brightness holding the
calibration parameters fixed, and estimation of the calibration
parameters holding the sky brightness model fixed
\citep{SELFCAL_LECTURE}.  For the first part, we use the combination
of the Wide-band \AWP\ \citep{WB-AWP} and the Multi-term
Multi-Frequency Synthesis algorithms \citep{MT-MFS}.  Hence in this
section, we will concentrate on the second part: estimating the
calibration parameters (pointing errors) while holding the sky
brightness fixed.

In Eq.~\ref{Eq:DCHIDD} the evaluation of the residual vector $R$, and
in this case also $\nabla \A_{ij}$, can be expensive. Both $R_{ij}$
and $\nabla_{\!\!i} \A_{ij}$ in Eq.~\ref{Eq:DCHIDD} involves
evaluation of $\A_{ij}$ and
$\left[\frac{\partial \A_i}{\partial
    a^{*}_{\!i}}  \ostar \A^{*}_j\right]$ at each iteration of the minimization
algorithm.  For an array with $N_{ant}$ antennas, the computational
cost of these evaluations is
${\it O}\left(4\times 2N_\A^2\log(N_\A) \times 2N^2_{ant}\right)$,
where $N_\A$ is the size in pixels of the quantized representation of
the elements of $\A$.
This cost can be
prohibitive for modern telescopes with $N_{ant}$ and $N_\A$ in the
range of few$\times 10^{2-3}$ antenna elements and $10^{5-7}$ pixels
respectively. Thus treating the PB as being completely unknown is not
viable computationally nor can we expect it to be well-conditioned. Hence 
a form for the PB with less degrees of freedom is necessary.

The mechanical antenna pointing errors can be represented with fewer
degrees of freedom; as a phase gradient across the antenna aperture
illumination pattern.  Purely mechanical fractional pointing errors
are also the same for both polarization and at all frequencies.
Aperture-plane solvers for the pointing errors can therefore easily
benefit from the instantaneous continuum sensitivity of modern
wide-band receivers.  For simplicity and to facilitate analysis, we
model the main-lobe of the EFP for antenna $i$ as a gaussian of the
form $e^{-\left(l-l_i\right)^2\alpha^2}$ where $l$ is a direction on
the sky with respect to the pointing direction, $l_i$ is the pointing
error and $\alpha$ is $2^{-1/2}$ times the inverse of the standard deviation.
It can be replaced with a more accurate function for
the antenna PB {\it without} the pointing errors in the final results.
$\A_{ij}$ (the Fourier transform of the PB) can then be
expressed as
\begin{equation}
\label{Eq:AIJPOINTING}
\A_{ij}=\A^{\!\circ}_{ij}\left[e^{-\frac{\left(l_i-l_j\right)^2\alpha^2}{2}}\right]e^{\iota\pi u\left(l_i+l_j\right)}
\end{equation}
and Eq.~\ref{Eq:DA} for
$a^{*}_{\!i}=l_i$ becomes
\begin{equation}
\label{Eq:DAIJPOINTING}
\nabla_{\!\!i} \A_{ij} = \frac{\partial \A_{ij}}{\partial l_i} =
\A_{ij}\left[\left(l_j-l_i\right)\alpha^2 +\iota \pi u\right]
\end{equation}
\BF{where $u$ is the Fourier-conjugate variable for $l$ (in units of
  wavelengths and radians respectively).}  The
equations above are written for one dimension only  for clarity and
can be trivially extended for the other dimension and for the
heterogeneous array case. $l_i$ and $l_j$ are 
the antenna pointing errors for antennas $i$ and $j$.
$\A^{\!\circ}_{ij}$ is the pre-computed version of $\A$ and includes
all PB effects known {\it a priori} -- but
{\it not} the mechanical antenna pointing errors -- like polarization
squint, off-axis polarization effects, effects of antenna blockages
and feeds, rotation with PA, dependence on frequency, etc.  Using
\AWP\ to apply $\A_{ij}$ and $\nabla_{\!\!i} \A_{ij}$, both $R_{ij}$
and $\partial{\chi^2}/\partial{a^{*}_{\!i}}$ can be computed at full
{\it continuum sensitivity} by integration across time, frequency and
polarization without loss of accuracy.  The term inside the square
brackets in Eq.~\ref{Eq:AIJPOINTING} is the reduction in the amplitude
because of the decorrelation of the measured visibilities due to the
pointing errors at each antenna of the baseline.  It is of order unity
for small values of $(l_i - l_j)$ -- the difference in the pointing
errors at the two antennas -- and may be ignored (for a typical
  maximum value for $(l_i-l_j)$ of order $1-2\%$ of the width of the antenna PB, this term 
  constitutes an error of less than $0.1\%$ in the amplitude).
\begin{algorithm}[H]
\caption{The Pointing Selfcal algorithm: estimation of pointing errors}
\label{Algo:PointingSelfcal}
\begin{algorithmic}[1]
%\Procedure{CH\textendash Election}{}
\State Pre-compute $V^{M^\circ} = \F\left[ I^M(\vec{s})\right]$
\State Pre-compute $\A^\circ_{ij}(t,\nu)$ for all required $i$, $j$, $t$ and $\nu$
\For{all data}
  \For {all $t$ and $\nu$ in the interval ($\tau_{sol}$,$\Delta \nu_{sol}$)}
       \For {Iteration $n$}
          \State {\tt Chi[] = 0.0; dChi[] =  0.0;}
            \For {all $i$}
            \For {all $j$, $j \neq i$}
              \State 
              \begin{varwidth}[t]{\linewidth} 
                \par Compute $\A_{ij}(t,\nu)$ and $\nabla_{\!i}  \A_{ij}(t,\nu)$\par
                \hskip\algorithmicindent
                (Eqs.~\ref{Eq:AIJPOINTING}, \ref{Eq:DAIJPOINTING})
                \end{varwidth}
              \State 
                 Use \AWP\ algorithm to compute:
                   \myindent{3.2}$V^M_{ij}= \A_{ij}\star V^{M^\circ}$ 
                   \myindent{3.2}$\nabla_{\!i} V^M_{ij}= \nabla_{\!i}\A_{ij} \star V^{M^\circ}$
                % \begin{varwidth}[t]{\linewidth} 
                %   \par Use the A-Projection algorithm to compute\par 
                %   \hskip\algorithmicindent  $V^M_{ij}= \A_{ij}\ostar
                %   V^{M^\circ}$ and
                %    $\nabla_{\!i} V^M_{ij}= \nabla_{\!i}\A_{ij} \ostar V^{M^\circ}$
                % \end{varwidth}
              \State Compute $R_{ij} =  V^{Obs}_{ij}(t,\nu) - V^M_{ij}(t,\nu)$
              \State 
                  Accumulate 
                    \myindent{2.9}{\tt Chi[j]=Chi[j]+}$R^{\dag}_{ij}W_{ij} R_{ij}$
                    \myindent{2.9}{\tt dChi[j]=dChi[j]+}$\Re\left(R_{ij}^{\dag}W_{ij}\left[\nabla_{\!i}V^M_{ij}\right]\right)$
                  % \begin{varwidth}[t]{\linewidth} 
                  %    \par Accumulate \par
                  % \hskip\algorithmicindent
                  %   {\tt Chi[j] =  Chi[j] + }$R^{*}_{ij}R_{ij}$
                  %   {\tt dChi[j] =  dChi[j] + }$R_{ij} + \nabla_{\!i} V^M_{ij}$
                  % \end{varwidth}
            \EndFor
            \If{$Terminate(${\tt Chi[],dChi[]},$n)$} 
            \label{TerminateStep}
              \State break
            \EndIf
            \State Update $\BARN{l_i}{n} = 
            \BARN{l_i}{n-1} +f\left({\tt dChi[j]}\right)$
            \label{UpdateStep}
            \EndFor
         \EndFor
\EndFor
\State Save all $l_i$ for solution intervals ($\tau_{sol}$, $\Delta \nu_{sol}$)
\EndFor
%\EndProcedure
\end{algorithmic}
\end{algorithm}
To predict the model data at each iteration, modified $\A$
as in Eq.~\ref{Eq:AIJPOINTING} is used with the \AWP\ algorithm to
compute the model data, including the effects of the antenna pointing
errors and subtracted from $V_{ij}^{Obs}$ to compute $R_{ij}$.
Similarly, modified $\A$ as in Eq.~\ref{Eq:DAIJPOINTING} is
used to compute
$\BARN{\nabla_{\!\!i}\A_{ij}}{n} \star V^{M^\circ}$.  See
Algorithm~\ref{Algo:PointingSelfcal} for the various computational
steps and the nesting of the loops involved.  The functions
$Terminate()$ and $f()$ in steps \ref{TerminateStep} and
\ref{UpdateStep} depend on the details of the minimization algorithm.
%%%%%%%%%%%%%%%%%%%%%%%%%
\section{Results}
\label{Sec:Results}
\subsection{Simulations}
To verify the numerical correctness and performance of the PSC
algorithm, we simulated data for the Karl G. Jansky Very Large Array
(VLA) \citep{THE-EVLA} which
included the effects of the time-varying pointing errors at each
antenna.  The simulation was for an L-Band observation and used a sky
model derived from the NVSS source list.  The total integrated flux in
the FoV, including the first sidelobe of the PB, was 1.5~Jy
distributed across the beam.  For numerical accuracy, the data for the
sky model was predicted using direct Fourier transform and a model for
the antenna PB with pointing offsets.  The pointing offsets for the
antennas were uniformly distributed between $\pm20\arcsec$, which
corresponds to $\sim\pm1\%$ of the beam-width at 1.5~GHz.  Independent
pointing offset errors of $\pm5\arcsec$ were added for each antenna
as a function of time to
simulate short term time-varying offsets.  Finally, random noise corresponding to
a continuum sensitivity limit of $1~\mu Jy$/beam in eight hours of
observing with an integration time of 10 sec per sample was added to
the visibilities to simulate thermal noise.

\begin{figure}[ht!]
\begin{center}
\includegraphics[width=9cm]{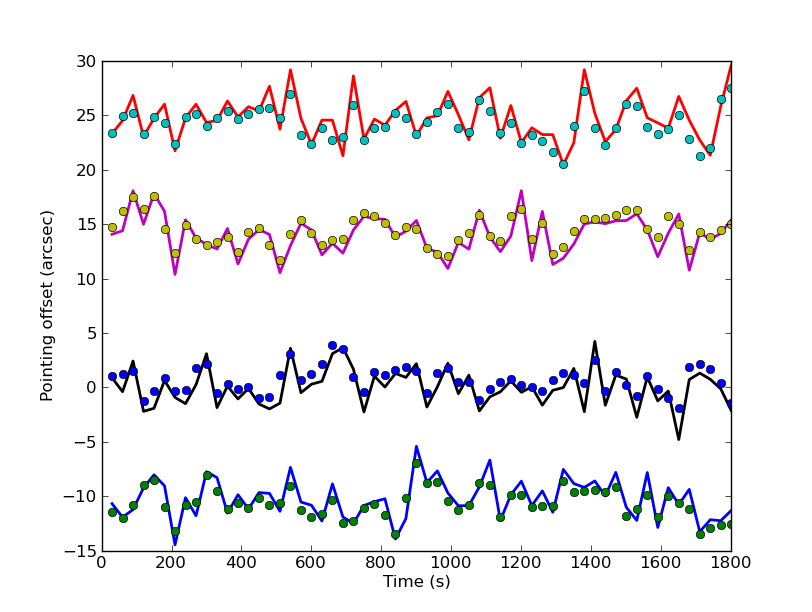}

\caption{\small The solutions from the Pointing Selfcal
  algorithm using simulated data to verify the algorithm.  The
  continuous lines show the pointing offsets as a function of time for
  four representative antennas while the filled-circles show the
  pointing offset solutions with 30 sec. solution interval.}
\label{FIGS:VERIFY}
\end{center}
\end{figure}

Figure~\ref{FIGS:VERIFY} shows the result of application of the PSC
algorithm to this simulated data.  The continuous curves show the
antenna pointing errors for 4 of the 27 antennas (for clarity) as a
function of time.  The over-plotted filled circles are the PSC
solutions with a solution interval of 30~sec.  The residuals per
baselines for solutions with 10~sec solution interval were consistent
with the thermal noise limit in the simulation
\cite[see][]{POINTING_SELFCAL}, verifying the basic numerical
correctness of the algorithm.

\begin{figure*}[ht!]
%\vskip -3.5in
%\hskip -0.25in
\centering
\hbox{
\includegraphics[width=9cm]{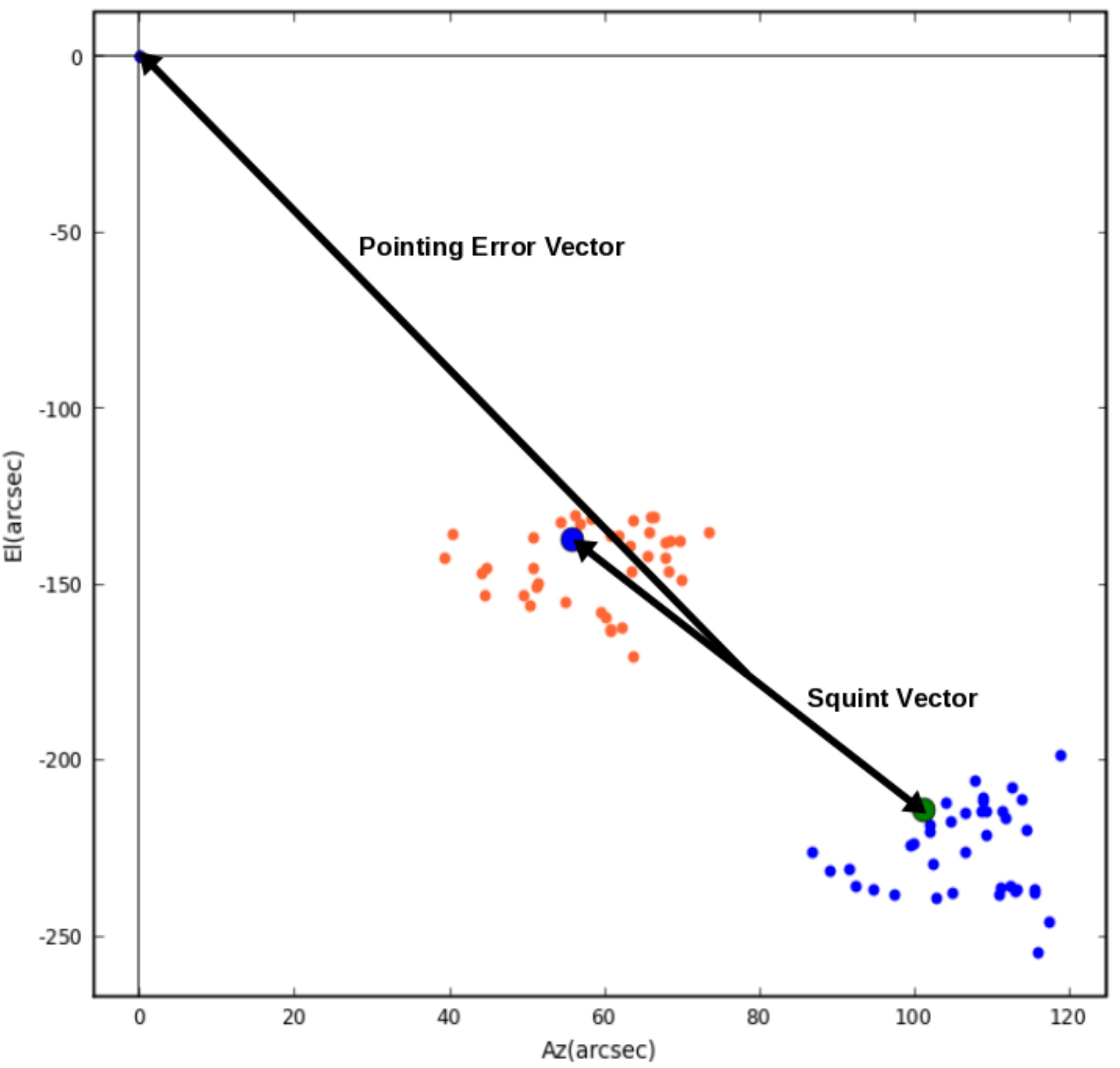}
\includegraphics[width=9cm]{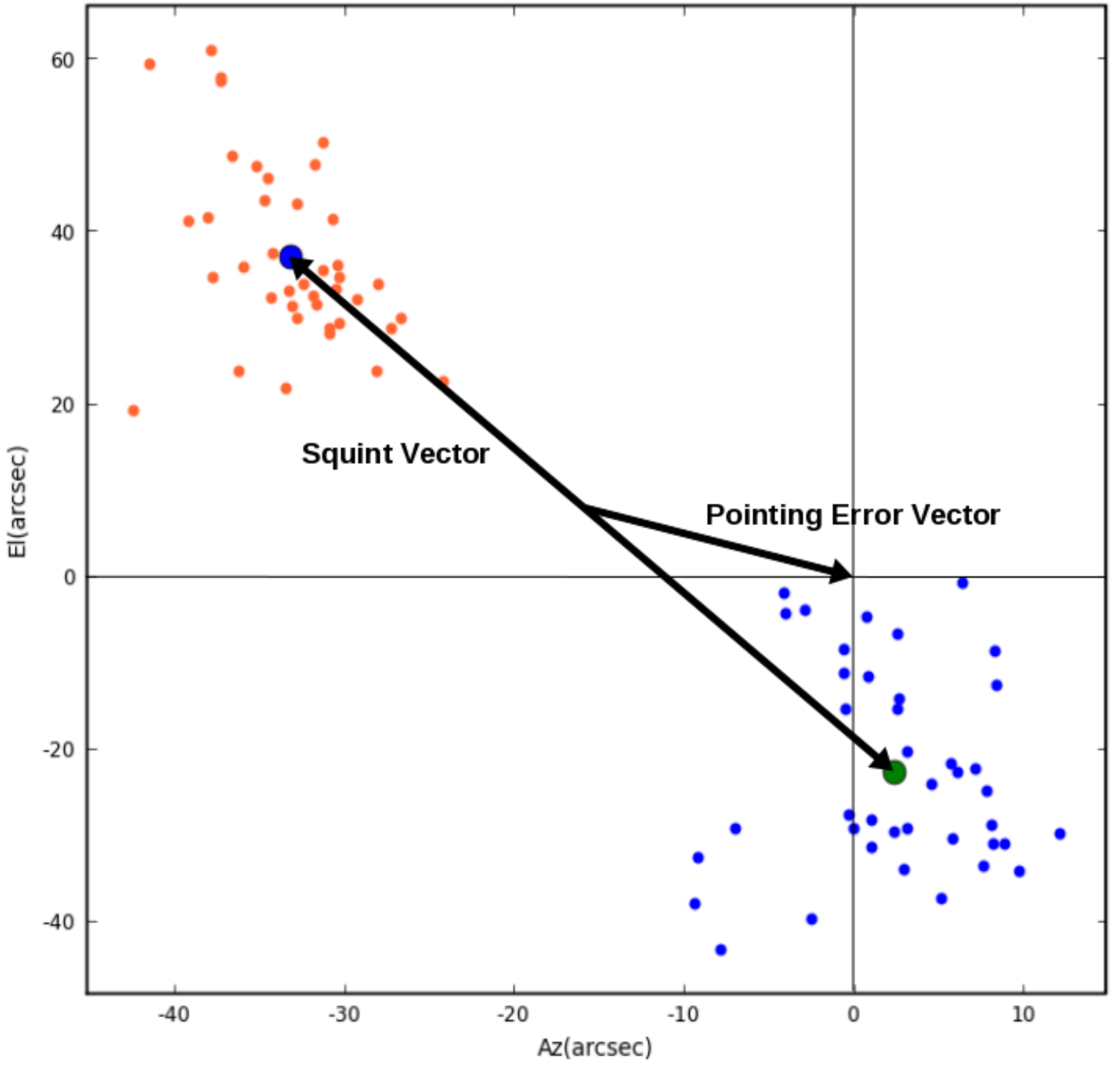}
}
\caption{\small The results from the application of the
  Pointing SelfCal algorithm to the VLA data.  Pointing error
  solutions along the elevation and azimuth axis for the R- and
  L-beams are shown with red and blue points.  The average separation
  between these set of points corresponds to the polarization squint
  of the VLA antennas due to the off-axis feed location.  The
  separation between the center of the squint vector and the origin
  corresponds to the antenna mechanical pointing error.}
\label{FIGS:OFFSETS_AZEL}
\end{figure*}

\begin{figure*}[ht!]
\begin{center}
\hbox{
\includegraphics[width=9cm]{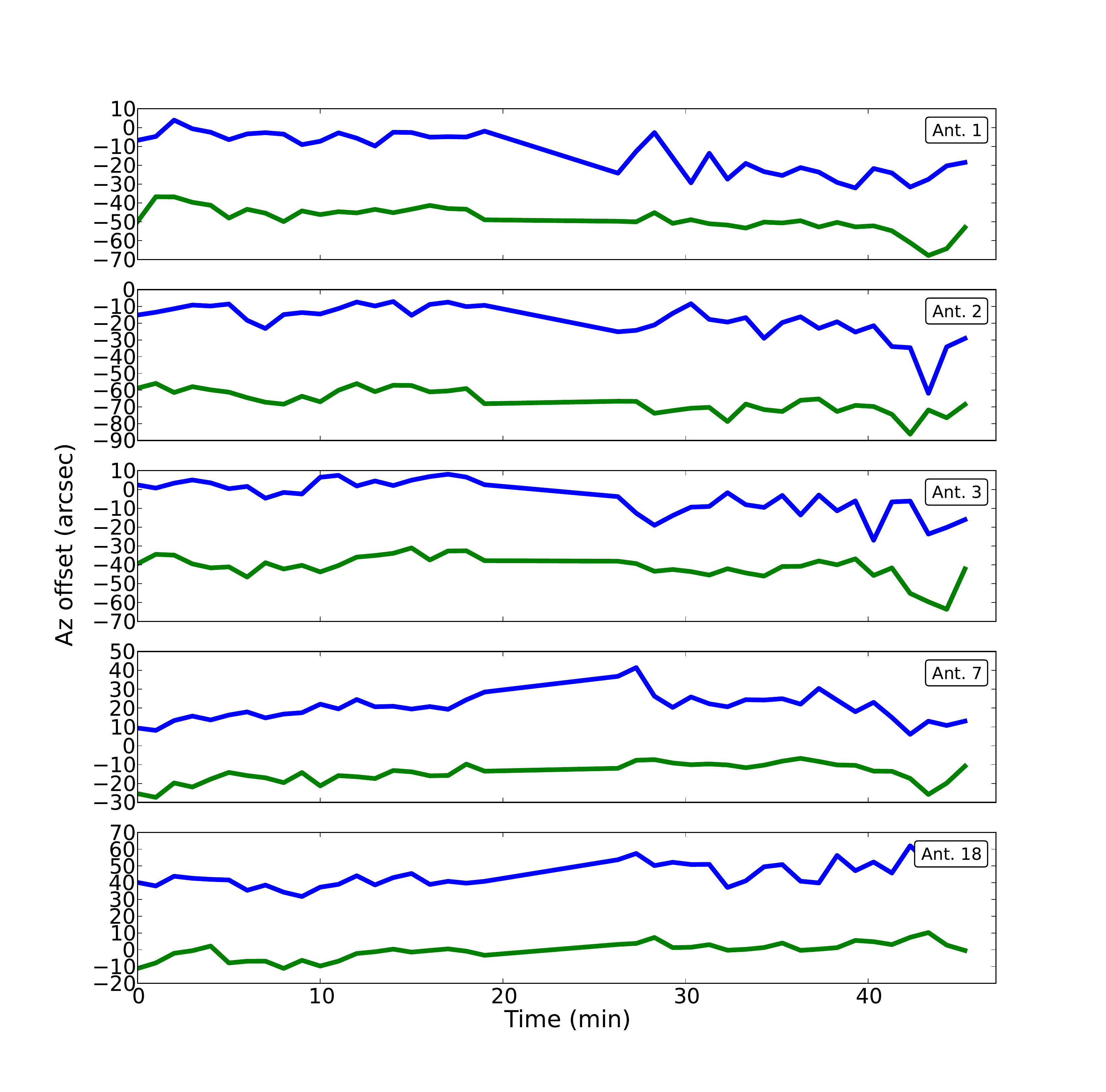}
\includegraphics[width=9cm]{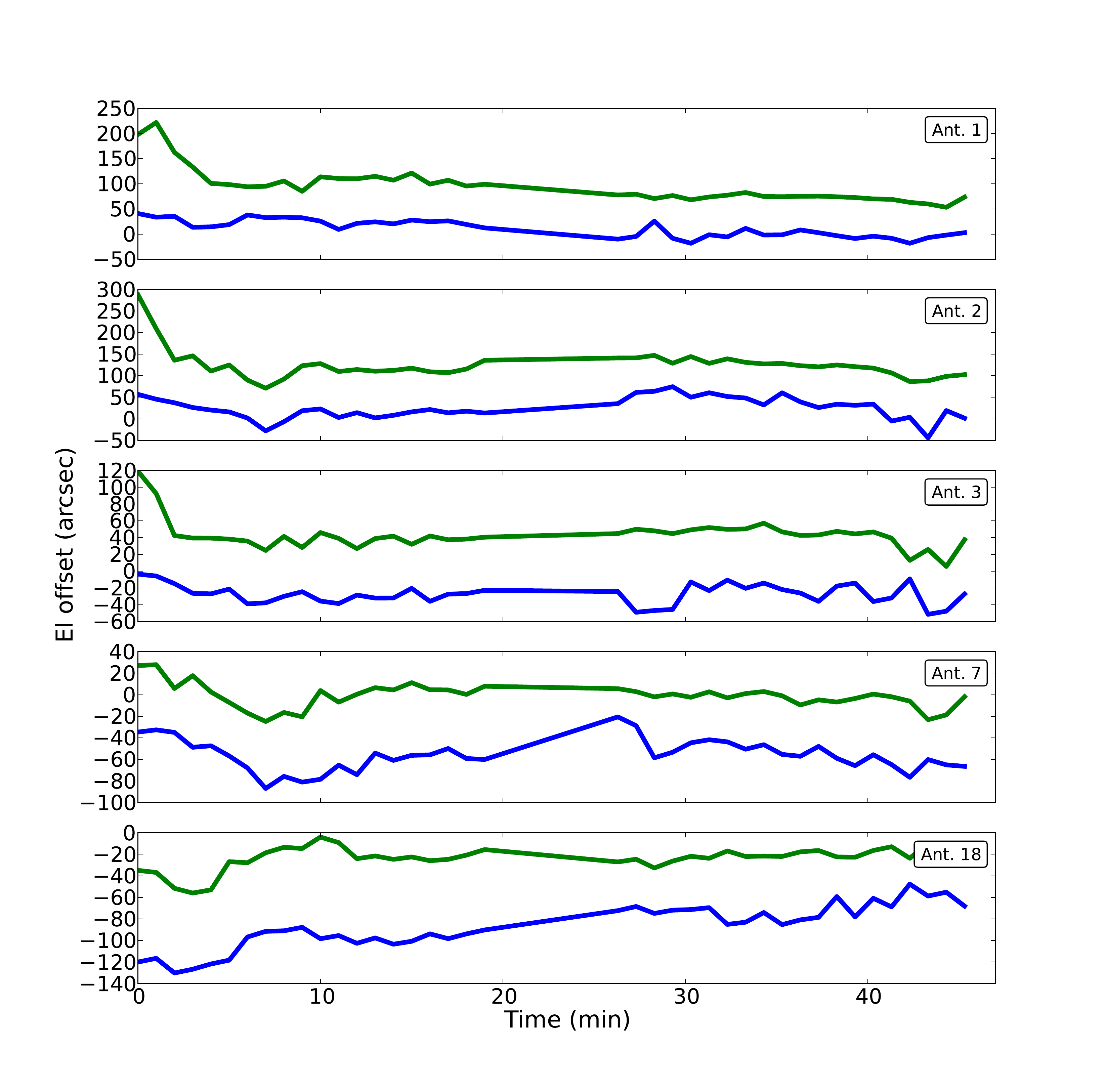}
}
\caption{\small The pointing offsets in the antenna azimuth
  (left) and elevation (right) axises as a function of time for
  several antennas.  The two curve in each panel correspond to the
  offsets for the R- and L-polarizations.  The separation between the
  two curves in each panel correspond to the azimuth and elevation
  components of the polarization squint for each antenna.}
\end{center}
\label{FIGS:AZVSTIME}
\end{figure*}

\subsection{Application to VLA data}
%Possible place for data: /home/tara/sanjay/CASATests/Pointing/ic2233_regression_data/

To test the algorithm with real data, we used a wide-band observation
of the IC2233 field with the VLA at L-Band.  This data was imaged
using about $\sim\!\!600$~MHz of bandwidth to generate the continuum
sky model used as an input for the PSC algorithm.  The VLA antenna
optics creates an offset between the parallel-hand PBs (polarization
squint) of $\sim\!\!6\%$ of the beam at any frequency, corresponding
to $\sim\!\!110\arcsec$ at 1.5~GHz.  To gain confidence in any antenna
pointing offset that the solutions may show, we set up the PSC
algorithm to solve for the offsets independently for both the
polarizations (R and L).  Difference in R- and L-solutions will then be
a measure of the polarization squint and the mean value of these
solutions ($[R_{offset} + L_{offset}]/2$) will be a measure of the
antenna mechanical pointing error (which should be the same for both
polarizations).  In the antenna Az-El plane, the separation between
the R and L solutions will then correspond to the length of the
polarization squint vector, while a vector from the center of the
squint vector to the origin of the Az-El plane would correspond to the
actual antenna mechanical pointing error vector.

The results on the Az-El plane for two of the antennas in the array
are shown in Fig.~\ref{FIGS:OFFSETS_AZEL}. The vector labeled ``Squint
Vector'' shows the separation between the median values of the two
offsets and is equal to $\sim\!\!120\arcsec$ and $105\arcsec$ for the
two antennas.  The vector labeled ``Pointing Error Vector'' from the
center of the Squint Vector to the origin is a measure of the antenna
mechanical pointing offsets, which has a magnitude of 3.5\arcmin and
0.5\arcmin~for these two antennas.  These are large systematic
pointing offsets, which were subsequently verified independently and
corrected in the telescope software.  As expected, after corrections,
particularly for the first antenna, an improved antenna sensitivity
was measured giving us a verification of the solutions and the sign
convention used in the software.

The Az- and El-offsets of the RR and LL beams as a function of time
from the nominal antenna pointing direction for a set of
representative antennas with solution-intervals of 5~min and 600~MHz
in time and frequency respectively are shown in
Fig.~\ref{FIGS:AZVSTIME}.  The PB model was derived using a geometric
optics simulator for the antenna illumination patterns which includes
the effects of aperture blockage, off-axis feed locations, and
illumination taper \citep{VLA_ILLUMINATION_PATTERN}.  The rotation
with PA and scaling of the PB with frequency was included in the model
using the \AWP\ algorithm.  The sky brightness for this observation is
dominated by two compact sources separated by $\sim\!\!25\arcsec$ with
other weaker sources spread across the FoV.  The scatter in the
pointing solutions is due to three factors:
\begin{enumerate}
\item The actual sky brightness is dominated by two strong sources and
  so the constraints on the perpendicular directions are weak.
\item Imperfections in the sky model
\item The limitations of the PB model used, particularly as a function
  of frequency.
\end{enumerate}

\begin{figure}[ht!]
% \vskip -3.5in
% \hskip -0.25in
\begin{center}
\includegraphics[width=9cm]{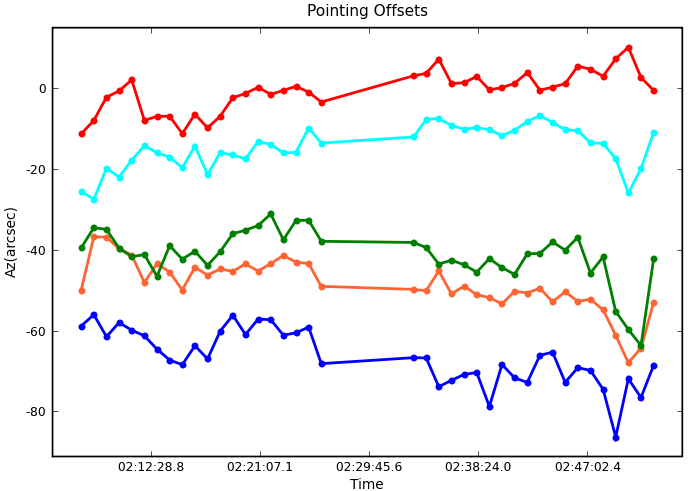}

\caption{\small The solved antenna pointing offsets in
  the azimuth axis with time for a few representative antennas of the
  VLA.}
\end{center}
\label{FIGS:IC2233-POINTINGOFFSETS}
\end{figure}

Analysis of independent holographic measurements
\citep{EVLA-HOLOGRAPHY} show that in addition to systematic deviations
from the expected value there are oscillations in the magnitude of the
squint vector as a function of frequency due to standing waves in the
antenna optics \citep{ASolver}, which were not included in the PB
model.  Some variations in the pointing offset solutions, including in
the length of the squint vector could therefore be also real. The
derived offsets as a function of time along the azimuth axis for a few
antennas in the array are shown in
Fig.~\ref{FIGS:IC2233-POINTINGOFFSETS}.  The solutions show
significant differences in the antenna pointing errors between the
antennas, as well as slow drifts over longer timescales
($\sim\!30~min$) -- both of which are expected based on the regular
pointing model measurements and anecdotal evidence from imaging
results.

%%%%%%%%%%%%%%%%%%%%%%%%%
\subsection{Noise budget}
\label{Sec:NoiseBudget}
In general, the stability and the error on the solved parameters
depend directly on the SNR in the measured data (the RHS of
Eq.~\ref{Eq:MEDD}) and on the number of free parameters.  It is
therefore important to devise algorithms that maximize data-SNR using
as few free parameters as possible.

The equivalent thermal noise that contribute to the variance in the
solution is reduced by a factor equal to the square root of the number
of statistically independent samples averaged.  The residuals in the
PSC solver are averaged at each iteration across all baselines with a
given antenna and for the duration of the solution intervals in time
and frequency.  Assuming that the pointing offsets at each antenna
are statistically independent and random in nature, the noise
contribution in the solver due to the thermal noise in the data is
reduced by a factor equal to
$\sqrt{\Delta\nu_{sol}\tau_{sol}\left(N_{ant}-1\right)}$ for solution
intervals of $\Delta\nu_{sol}$ and $\tau_{sol}$ in frequency and
time. On the other hand, the signal for the solver is the differential
apparent-flux with respect to the pointing offsets in the FoV.
Comparing this signal with the effective noise allows an estimate for
the magnitude and the distribution of the sky brightness required for
deriving pointing offset solutions.  For a homogeneous array
  case with identical antennas, the SNR per parameter for an
antenna-based pointing offsets solver is
\begin{equation}
SNR_{PSC} = \frac{\left|\nabla S\right|}{SEFD}~\sqrt{{\Delta \nu_{sol}\tau_{sol}}{\left(N_{ant}-1\right)}}
\end{equation}
where 
\begin{equation}
SEFD = \frac{2k_b T_{sys}}{(\eta_A \pi R^2)}\times 10^{26}~(Jy)\nonumber
\end{equation}
$k_b$ is the Boltzmann's constant.  $R$, $\eta_A$ and $T_{sys}$ are
the antenna aperture radius (in meters), efficiency and system
temperature (in Kelvin) respectively.  $\nabla S$ is the differential
apparent integrated flux in the FoV with respect to the antenna
pointing errors is given by:
\begin{equation}
\label{Eq:DELTA-S}
\nabla S= \int \left(\frac{\partial \E{}{}{}}{\partial l}
\otimes \E{}{*}{}\right)~I^M(\vec{s})~
    d\vec{s}~~~~(Jy)
\end{equation}
where $\E{}{}{}$ is the antenna far-field EFP\footnote{Recall that
  $\E{}{}{}$ is related to Eq.~\ref{Eq:ME} via
  $\MS{ij}{DD}{}=\E{i}{}{}\otimes\E{i}{*}{}$ and the Fourier transform
  of $\E{}{}{}$ is the DD equivalent of $\J$ in Eq.~\ref{Eq:MEDI}} and
$I^M$ is a model for the sky brightness distribution as a function of
the direction $\vec{s}$.  For the VLA at L-band, $N_{ant}=27$ and
$SEFD\approx\!358$~Jy ($T_{sys}=35$~K, $\eta_A=0.55$ and $R=12.5$~m).
The total apparent flux in the VLA FoV at L-band above the thermal
noise limit for a usable bandwidth of $800$~MHz and 1~min of
integration in time is estimated to be few$\times100$~mJy.  In the
absence of any other source of noise, solutions for antenna pointing
offsets should be possible at reasonably high accuracy on very short
timescales. Some numerical experiments suggest that reliable solutions
are possible at several minutes timescale due to addition numerical
noise, e.g. due to gridding, rotation of $\A^M$ with PA, inaccuracies
in $I^M$ and $\A^M$, etc.

\begin{figure}
\begin{center}
\includegraphics[width=8cm]{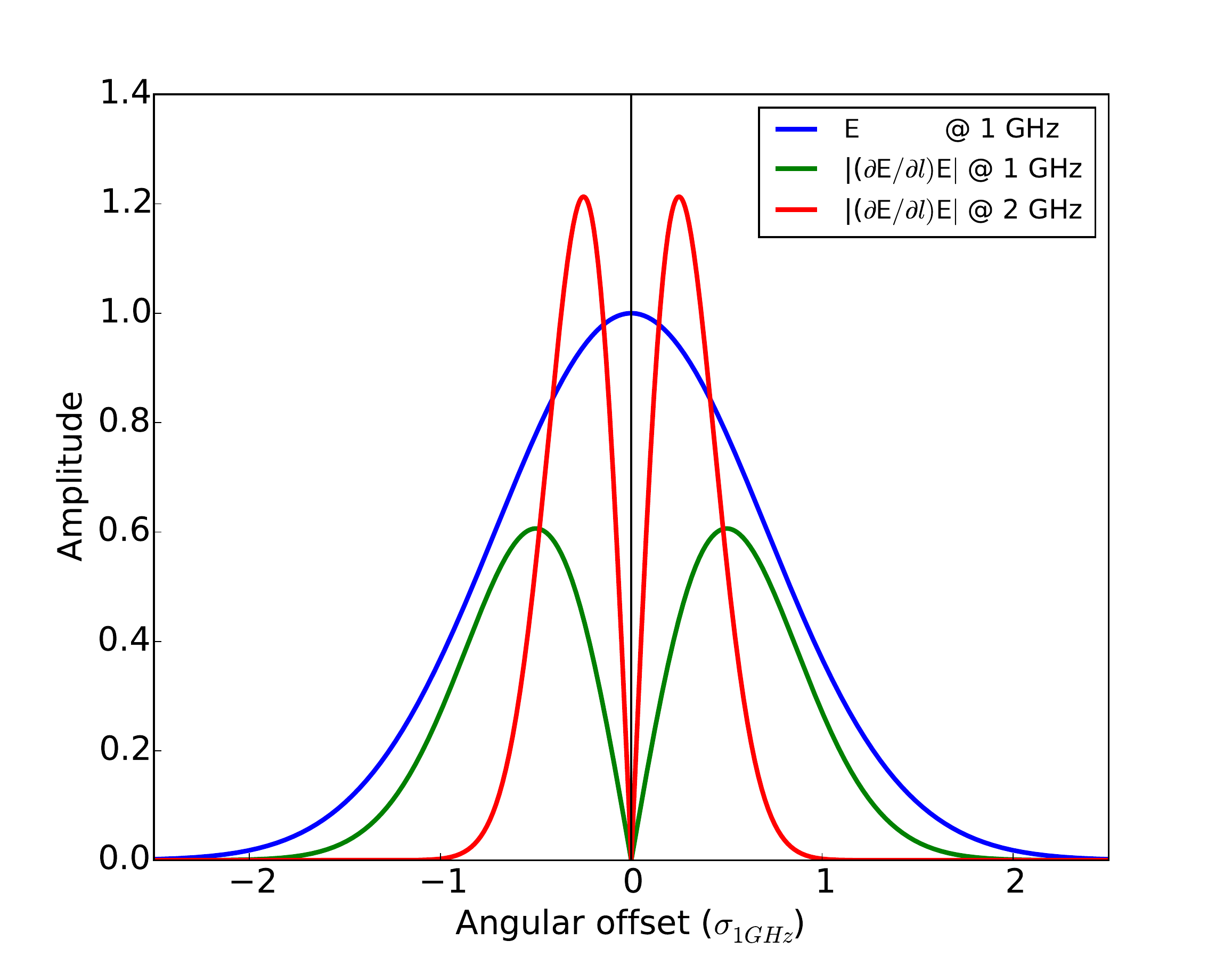}

\caption{\small The curves show slices across the
  $\left|\E{}{}{(\nu)}\partial\E{}{}{(\nu)}/\partial l\right|$ function at the edges of a 1-GHz
  wide band (in green and red). The main-lobe of $\E{}{}{}$ at 1~GHz is
  approximated as a gaussian of width $\sigma_{1GHz}$ (in blue).}
\end{center}
\label{FIGS:dE}
\end{figure}

Analysis of Eq.~\ref{Eq:DELTA-S} offers a few useful thumb-rules.  The
function
$\left|\frac{\partial \E{}{}{}}{\partial l}\otimes\E{}{*}{}\right|$ is
a double-hump shaped curve with a null at the location of the peak of
$\E{}{}{}$ (see Fig.~\ref{FIGS:dE}).  Therefore, while the sky
brightness {\it at} the center of the PB does not contribute signal
for the solver, the contribution of the flux around the center
increases with the magnitude of the pointing error.  This function
peaks around the half-power-point of the PB.  The flux in this region
of the PB therefore contributes the maximum signal.  The PSC algorithm
therefore works well for fields where the sky brightness distribution
is spread across the PB.  This is almost always the case at low
frequencies ($\lessapprox$few~GHz) and for mosaic imaging which is
typically the observing mode at high frequencies.  Note that the
antenna pointing errors also adversely affect imaging performance only
for fields with significant flux away from the pointing center.  
  The area under this curve is independent of frequency (changes in
  ${\partial \E{}{}{}}/{\partial l}$ and $\E{}{}{}$ compensates for
  each other as a function of frequency). Since the mechanical
  pointing errors are independent of frequency and on an average the
  low-frequency radio flux varies as $\nu^{-0.7}$, the signal for the
  pointing solver will increase with bandwidth, particularly at lower
  radio frequencies.  More precise estimates for the SNR and solution
  timescales will require simulations using models for $\E{}{}{}$, 
  the sky brightness and its spectral index distribution.

%%%%%%%%%%%%%%%%%%%%%%%%%
\section{Discussions}
\label{Sec:Discussions}
%%%%%%%%%%%%%%%%%%%%%%%%%
\subsection{Run-time performance analysis}
\label{Sec:RunTime}
%%%%%%%%%%%%%%%%%%%%%%%%%
The run-time cost of PSC has two components: computation of $\A_{ij}$
and $\nabla_{\!\!i} \A_{ij}$ and their application to $V_{ij}^{Obs}$
via \AWP.  Using Eqs.~\ref{Eq:AIJPOINTING} and \ref{Eq:DAIJPOINTING}
with a precomputed $\A^{\circ}_{ij}$, the cost of evaluating $\A_{ij}$
and $\nabla_{\!i}\A_{ij}$ at each iteration becomes relatively
insignificant and the total run-time cost is dominated by the cost of
their application to $V_{ij}^{Obs}$.  For a support size of $N_{sup}$
for $\A_{ij}$, this cost scales as $N_{vis}N^2_{sup}$, where $N_{vis}$
is the total number of data samples within the solution interval.
Since the computations for each data sample is independent, this cost
reduces linearly with parallelization by partitioning $N_{vis}$ across
multiple computing cores.  As a test case using a single CPU running
at 1.2~GHz clock, each iteration of the PSC took about 1~sec per
frequency channel for the VLA.  Using all the 16 computing cores
available, the run-time was reduced to $\sim70$~millisec per channel,
with the total run-time for convergence of $\sim2$~min per solution.

Multi-threaded gridders (e.g., \cite{MULTITHREADED-GRIDDER}) which can
benefit from a much larger number of compute-cores available on
massively parallel hardware like the modern GPUs may help in reducing
the run-time cost by a large factor.  For arrays that are non-coplanar
for long-integration observations, the run-time cost of PSC increases
with the W-term \citep{WProjection_IEEE}.  However, for short solution interval the run-time
cost may be largely independent of the W-term -- by using the fact
that arrays are instantaneously co-planar \citep{W-SNAPSHOT}.  
However, more work is needed in these areas to determine the optimal computing 
architecture for PSC.

\subsection{Use in real-time-calibration}
Some modern-era radio telescopes assume that a Local Sky Model is
available and use it to determine calibration parameters in real time
without resorting to iteration over the model (see {\em e.g.}
\cite{2012CRPhy..13...28T}).  With some additional cost in computing,
our algorithm for estimation of the pointing errors could be used to
track any antenna pointing errors in real time and possibly correct
the pointing errors in real-time.

\subsection{Solving for PB shape}
The PSC algorithm described above solves for the tip-tilt of the
antenna by solving for a phase gradient across the antenna aperture.
This does not however alter the Hermitian nature of an ideal aperture
illumination pattern and therefore does not alter the predicted shape
of the antenna PB.  In practice, change in the shape of the antenna PB
arises due to a variety of reasons including de-focus and astigmatism
in the antenna optics, distortion of the main reflector with
elevation, misplaced or misaligned elements in the antenna optics, and
projection effects in aperture-array elements.  Many of these terms
result in a non-Hermitian aperture illumination pattern which, in
addition to shape distortions also leads to complex-value PB.  While
pointing errors constitute the dominant direction-dependent error,
errors due to the shape of the PB are also significant for the
sensitivity offered by all modern radio telescopes.  Its calibration
is therefore also necessary.

The low-order A-Solver approach \citep{ASolver} offers a method for a
closed-loop Shape Selfcal (SSC) similar to the PSC algorithm.  The
A-Solver uses a {\it physically} motivated parametrized model of the
antenna structure in a geometric optics (GO) simulator
\citep{VLA_ILLUMINATION_PATTERN} for the antenna illumination pattern
(AIP).  Starting with reasonable values for these physical parameters,
the A-Solver solves for these parameters by minimizing the difference
between the predicted and holographically measured AIP.  Since the
model AIP needs to be predicted in the optimization iterations, it is
important to use a simulator with a reasonably short run-time.
Full-EM simulators are typically quite expensive (both, in capital and
run-time costs). GO simulators on the other hand are less
computationally complex and when used in the A-Solver, capture the
dominant electromagnetic effects in the resulting optimized model.
\cite{ASolver} show that this approach captures otherwise difficult to
model effects like the effect of Standing Waves in the antenna optics
and other higher order phase terms across the antenna aperture.  These
higher order phase terms severely affect the off-axis polarization
leakage patterns and the A-Solver approach offers an effective method
that can enable noise-limited full-Stokes imaging (not just Stoke-I
imaging).  The error analysis in Sec.\ref{Sec:NoiseBudget} above
suggests that the use of a sky-model instead of the holographic
measurements may be possible (from an SNR point of view) in the
A-Solver for a closed-loop SSC algorithm.  Note however that for many
antenna arrays, shape changes are smooth and gradual and it may be
sufficient to first derive the AIP model using holographic
measurements and include the temporal evolution of the derived
parameters separately.  For the aperture-array elements, where these
evolutions are more severe and faster, a closed-loop SSC may be
required, and possible, using a model for the sky-brightness
distribution (e.g. a pre-determined Global Sky Model).

%%%%%%%%%%%%%%%%%%%%%%%%%
\section{Conclusions}
\label{Sec:Conclusions}
In this paper we present the mathematical framework for calibration of
direction-dependent effects (DDE) not known {\it a priori}.  We also
include a brief overview of the direction-independent (DI) calibration
and the mathematical formulation of the existing DI SelfCal algorithm.

As an example of a DD calibration algorithm, we present the Pointing
Selfcal (PSC) algorithm.  As in DI SelfCal, given a model of the sky
brightness distribution, the pointing offset vector per antenna is
solved by iteratively minimizing the residual vector with respect to
the antenna-based point offsets.  We verified the PSC solver, first by
applying it to simulated wide-band data for the VLA at L-band and
show that the pointing offset vector is recovered correctly.  We then
also apply the PSC algorithm to on-sky data from the VLA at L-band.
The VLA antenna optics has a polarization squint which results in an
angular separation between the right- and left-circular polarization
beams.  To verify the PSC algorithm, the solver was set-up to solve
for the pointing offset vector separately for the two polarizations.
In the antenna Az-El plane, the difference between the pointing offset
vectors for the two polarizations is a measure of the squint vector and
the separation of the center of squint vector from the origin gives a
measure of the mechanical antenna pointing offset.  We verified that the PSC
solver indeed recovers the average squint vector, though
antenna-to-antenna variations were also large and significant.  Some
of the antennas had large mechanical pointing offsets, which were
subsequently verified via independent measurement of the expected
improvement in the antenna gain after correcting for them in the
telescope software.

We also discuss the noise budget for the PSC algorithm.  Analysis of
the signal-to-noise ratio (SNR) available for the PSC solver as a
function of the wide-band sky brightness distribution and telescope
parameters leads to the following conclusions:
\begin{enumerate}
\item The PSC algorithm is optimal in utilizing the SNR due to the sky
  brightness distribution in the {\it entire} antenna field of view, 
  rather than, for example, a few bright sources.
\item Antenna pointing offsets can be solved-for at high significance
  with the instantaneous sensitivity of most modern radio
  interferometric telescopes with wide-band
  receivers and typical sky brightness distributions.
\item While the sky brightness {\it at} the center of the PB does not
  contribute signal for the PSC solver, the contribution from around
  the center increases with the magnitude of the pointing offsets.
  This signal also peaks around the half-power points of the
  PB.  The PSC algorithm therefore works well for observations where
  the sky brightness is distributed across the FoV.  The degradation
  in the imaging performance due to the antenna pointing errors is also
  more significant for such observations.  Emission spread across the
  FoV is typical at frequencies below a few GHz and at much higher
  frequencies where mosaic imaging of emission much larger than the
  antenna PB is often necessary.
\item We expect PSC to scale well in a parallel computing
  environment.  Simple parallelization by data-partitioning is
  possible and efficient.  Reduction in the run time by large factors
  using multi-threaded re-samplers \citep{MULTITHREADED-GRIDDER}
  deployed on massively parallel hardware remains a possibility,
  though more work is needed in this area to arrive at an optimal
  computing architecture.
\item The PSC is typically set-up for relatively short solution
  interval in time.  For low-frequency observations, the increase in
  the run time due to the w-term may be mitigated by treating the array as
  co-planar for each solution interval \citep{W-SNAPSHOT}.
\end{enumerate}

The mathematical framework for DD calibration presented here can be
extended for a Shape SelfCal (SSC) algorithm to account for the change
in the shape of the aperture using the low-order A-Solver approach.
In the A-Solver approach, the parameters describing the physical
structure of the antenna are determined using a geometric optics
predictor for the antenna aperture illumination pattern (AIP) and
holographic measurements of the AIP \citep{ASolver}.  Based on our
estimate of the SNR typically available, it may be possible to develop
an SSC algorithm using a model of the sky brightness distribution.
More work is required, and in progress, in this area.

Finally, our algorithm could be used in the real-time calibration
systems of modern-era radio telescopes where a Local Sky Model is
assumed to be available.  This will allow tracking of any antenna
pointing errors in real time and possibly real-time corrections.
\begin{acknowledgements}
  This work was done using the R\&D branch of the CASA code base.
  \BF{We wish to thank the referee, Daniel Mitchell, for very useful
    comments and pointing out some corrections in the equations and
    their interpretation.}  The National Radio Astronomy Observatory
  is a facility of the National Science Foundation operated under
  cooperative agreement by Associated Universities, Inc.
\end{acknowledgements}
\software{CASA\footnote{https://casa.nrao.edu}, Python, MatPlotLib}
\bibliography{bhatnagar08}

\end{document}